\documentclass[reprint,superscriptaddress,amsmath,amssymb,amsfonts,aps,floatfix]{revtex4-2}

\usepackage{natbib}
\usepackage{graphicx,color,rotating}
\usepackage[latin1]{inputenc}
\usepackage{textcomp}
\usepackage{dcolumn}
\usepackage{bm}     
\usepackage{upgreek}
\usepackage{array} 
\newcolumntype{L}[1]{>{\raggedright\let\newline\\\arraybackslash\hspace{0pt}}m{#1}}
\newcolumntype{C}[1]{>{\centering\let\newline\\\arraybackslash\hspace{0pt}}m{#1}}
\newcolumntype{R}[1]{>{\raggedleft\let\newline\\\arraybackslash\hspace{0pt}}m{#1}}

\usepackage{hyperref}					
\usepackage{xcolor}						
\hypersetup{			     			
    colorlinks,
    allcolors={blue}
}
\newcommand{\qmarks}[1]{{``#1''}}
\newcommand{\qmarkstt}[1]{{``\texttt{#1}''}}

\newcommand{\upspace}{\rule{0ex}{2.5ex}}

\newcommand{\notop}{{{}_{}}}

\newcommand{\mr}[1]{\ensuremath{\mathrm{#1}}}
\newcommand{\myvec}[1]{\bm{#1}}
\newcommand{\ee}{\mathrm{e}}
\newcommand{\ii}{\mathrm{i}}
\newcommand{\dm}{\mathrm{d}}

\newcommand{\avr}[1]{\big\langle #1 \big\rangle}

\DeclareMathOperator{\re}{Re}

\newcommand{\ve}{\varepsilon}
\newcommand{\veO}{\ve_0}

\newcommand{\pp}{\partial}

\newcommand{\nablabf}{\boldsymbol{\nabla}}

\newcommand{\Lapl}{\nabla^2}

\newcommand{\grad}{\nablabf}

\renewcommand{\div}{\nablabf\cdot}

\newcommand{\etal}{\textit{et~al.\ }}

\newcommand{\Cnn}{C^\notop}
\newcommand{\DDD}{\myvec{D}}

\newcommand{\een}{\myvec{e}}

\newcommand{\FFF}{\myvec{F}}

\newcommand{\FFFrad}{\myvec{F}^\mathrm{rad}}

\newcommand{\Frad}{F^{\mathrm{rad}}}

\newcommand{\Hsl}{{H^\notop_\mr{sl}}}	
\newcommand{\Hpz}{{H^\notop_\mr{pz}}}	
\newcommand{\Hgr}{{H^\notop_\mr{gr}}}	
\newcommand{\Hgl}{{H^\notop_\mr{gl}}}

\newcommand{\Hfl}{H^\notop_\mr{fl}}

\newcommand{\kc}{k_\mathrm{c}}

\newcommand{\ks}{k_\mathrm{s}}

\newcommand{\Lfl}{L^\notop_\mr{fl}}
\newcommand{\Lsl}{L^\notop_\mr{sl}}
\newcommand{\Lpz}{L^\notop_\mr{pz}}
\newcommand{\upLpz}{{L_\mr{pz}}}
\newcommand{\Lcl}{L^\notop_\mr{cl}}

\newcommand{\nnn}{\myvec{n}}

\newcommand{\Pin}{P_\mathrm{in}}

\newcommand{\rrr}{\myvec{r}}

\newcommand{\uuu}{\myvec{u}}
\newcommand{\uun}{\myvec{u}}
\newcommand{\unn}{{u}}
\newcommand{\VVV}{\myvec{V}}

\newcommand{\vvv}{\myvec{v}}
\newcommand{\vvn}{\myvec{v}}
\newcommand{\vnn}{{v}}

\newcommand{\Wgr}{{W^\notop_\mr{gr}}}
\newcommand{\Wpz}{{W^\notop_\mr{pz}}}
\newcommand{\Wsl}{{W^\notop_\mr{sl}}}
\newcommand{\Wfl}{{W^\notop_\mr{fl}}}
\newcommand{\Wbin}{{W^\notop_\mr{bin}}}
\newcommand{\Lbin}{{L^\notop_\mr{bin}}}
\newcommand{\Nbin}{{N^\notop_\mr{bin}}}

\newcommand{\zerovec}{\boldsymbol{0}}

\newcommand{\tac}{t^\notop_\mr{ac}}

\newcommand{\calC}{\mathcal{C}}

\newcommand{\cO}{c_0}

\newcommand{\Eac}{E_\mathrm{ac}}

\newcommand{\Eacfl}{E_\mr{ac}^\mr{fl}}

\newcommand{\FFFdrag}{\FFF^{\mathrm{drag}_{}}}

\newcommand{\Dy}{\Delta y}

\newcommand{\etaBO}{\eta^\mathrm{b}_0}

\newcommand{\etaOb}{\eta^\mathrm{b}_0}

\newcommand{\etaO}{\eta_0}

\newcommand{\Gamsl}{\Gamma_\mathrm{sl}}

\newcommand{\vph}{\varphi}
\newcommand{\vphn}{\varphi^\notop}

\newcommand{\rhosl}{\rho_\mr{sl}}

\newcommand{\fO}{f_0}

\newcommand{\fI}{f_1}

\newcommand{\fres}{f_\mathrm{res}}

\newcommand{\kO}{k_0}

\newcommand{\kapO}{\kappa_0}

\newcommand{\pI}{p_1}

\newcommand{\sa}{\text{sa}}
\newcommand{\asa}{\text{asa}}

\newcommand{\vvvI}{\vvv_1}

\newcommand{\rhoO}{\rho_0}

\newcommand{\rhoI}{\rho_1}

\newcommand{\SIC}{\textrm{C}}
\newcommand{\SICel}{^\circ\!\textrm{C}}
\newcommand{\SIcm}{\textrm{cm}}
\newcommand{\SIum}{\upmu\textrm{m}}

\newcommand{\SIMHz}{\textrm{MHz}}
\newcommand{\SIkHz}{\textrm{kHz}}
\newcommand{\SIJ}{\textrm{J}}

\newcommand{\SIJpmc}{\textrm{J}/\textrm{m$^3$}}

\newcommand{\SImL}{\textrm{mL}}
\newcommand{\SImuL}{\textrm{\textmu{}L}}
\newcommand{\SImuLpmin}{\SImuL\:\textrm{min}$^{-1}$}

\newcommand{\SIkg}{\textrm{kg}}
\newcommand{\SIkgm}{\textrm{kg}\:\textrm{m$^{-3}$}}

\newcommand{\SImin}{\textrm{min}}
\newcommand{\SIm}{\textrm{m}}

\newcommand{\SImm}{\textrm{mm}}
\newcommand{\SImum}{\textrm{\textmu{}m}}

\newcommand{\SIpN}{\textrm{pN}}

\newcommand{\SIkPa}{\textrm{kPa}}
\newcommand{\SIpTPa}{\textrm{TPa}^{-1}}

\newcommand{\SImPas}{\textrm{mPa}\:\textrm{s}}

\newcommand{\SIs}{\textrm{s}}

\newcommand{\SImps}{\SIm\,\SIs^{-1}}
\newcommand{\SImumps}{\SImum\,\SIs^{-1}}

\newcommand{\SIV}{\textrm{V}}

\newcommand{\SImW}{\textrm{mW}}

\newcommand{\nn}{\nonumber}
\newcommand{\beq}[1]{\begin{equation} \eqlab{#1}}
\newcommand{\eeq}{\end{equation}}
\newcommand{\bsub}{\begin{subequations}}
\newcommand{\esub}{\end{subequations}}
\def\bal#1\eal{\begin{align}#1\end{align}}
\def\balat#1#2\ealat{\begin{alignat}{#1} #2 \end{alignat}}
\def\bsubal#1 #2\esubal{\bsuba{#1}\begin{align}#2\end{align} \esuba}     
\def\bsubalat#1#2#3\esubalat{\bsuba{#1} \begin{alignat}{#2} #3 \end{alignat} \esuba}
\newcommand{\bsuba}[1]{\bsub \eqlab{#1}}
\newcommand{\esuba}{\esub}

\newcommand{\emiot}{\ee^{-\ii\omega t}}

\newcommand{\eqlab}[1]{\label{eq:#1}}
\renewcommand{\eqref}[1]{Eq.~(\ref{eq:#1})}
\newcommand{\eqnoref}[1]{(\ref{eq:#1})}

\newcommand{\eqsref}[2]{Eqs.~(\ref{eq:#1}) and~(\ref{eq:#2})}
\newcommand{\eqsnoref}[2]{(\ref{eq:#1}) and~(\ref{eq:#2})}

\newcommand{\figref}[1]{Fig.~\ref{fig:#1}}
\newcommand{\fignoref}[1]{\ref{fig:#1}}
\newcommand{\figsref}[2]{Figs.~\ref{fig:#1} and~\ref{fig:#2}}
\newcommand{\figlab}[1]{\label{fig:#1}}

\newcommand{\secref}[1]{Section~\ref{sec:#1}}

\newcommand{\seclab}[1]{\label{sec:#1}}
\newcommand{\tabref}[1]{Table~\ref{tab:#1}}

\newcommand{\tablab}[1]{\label{tab:#1}}

\newcommand{\vebf}{\bm{\ve}}

\newcommand{\sigmabf}{\bm{\sigma}}
\newcommand{\ppperp}{\partial^\notop_\perp}
\newcommand{\ppperpsqr}{\partial_\perp^2}

\newcommand{\sigmabfsl}{\bm{\sigma}^{{}}_\mr{sl}}

\newcommand{\cL}{c_\mathrm{lo}}

\newcommand{\cT}{c_\mathrm{tr}}

\newcommand{\uuuI}{\myvec{u}_1}

\newcommand{\fl}{\mathrm{fl}}

\newcommand{\pz}{\mathrm{pz}}
\renewcommand{\sl}{\mathrm{sl}}

\definecolor{darkgreen}{rgb}{0.00, 0.50, 0.00}
\definecolor{DARKGREEN}{rgb}{0.00, 0.50, 0.00}
\definecolor{RED}{rgb}{1.00, 0.00, 0.00}
\definecolor{GREEN}{rgb}{0.00, 1.00, 0.00}
\definecolor{BLUE}{rgb}{0.00, 0.00, 1.00}
\definecolor{MAGENTA}{rgb}{1.00, 0.00, 1.00}

\newcommand{\atsurface}{0}				

\newcommand{\vvvwall}{\VVV^\atsurface}				

\newcommand{\zs}{\zeta}

\newcommand{\pardiv}{\nablabf_{\parallel}^\notop\!\!\cdot}

\renewcommand{\perp}{\zeta}

\newcommand{\vwallperp}{\vwall_{1\perp}}
\newcommand{\vvvwallpar}{\vvvwall_{1\parallel}}

\newcommand{\vwall}{V^\atsurface}					

\newcommand{\ezs}{\een_\zs}

\begin{document}

\title{Increased throughput in
antisymmetrically actuated\\ acoustofluidic flow-through devices}

\author{Klara Andersson}
\email{klara.andersson@bme.lth.se}
\affiliation{Department of Biomedical Engineering, Lund University, Ole R\"{o}mers v\"{a}g 3, 22363 Lund, Sweden}

\author{Sazid Z. Hoque}
\email{shoque@iitg.ac.in}
\affiliation{Department of Physics, Technical University of Denmark,\\
DTU Physics Building 309, DK-2800 Kongens Lyngby, Denmark}
\affiliation{Department of Mechanical Engineering, Indian Institute of Technology Guwahati,\\
North Guwahati, Guwahati Assam-781039, India}

\author{Wei Qiu}
\email{wei.qiu@bme.lth.se}
\affiliation{Department of Biomedical Engineering, Lund University, Ole R\"{o}mers v\"{a}g 3, 22363 Lund, Sweden}

\author{Andreas Lenshof}
\email{andreas.lenshof@bme.lth.se}
\affiliation{Department of Biomedical Engineering, Lund University, Ole R\"{o}mers v\"{a}g 3, 22363 Lund, Sweden}

\author{Pelle Ohlsson}
\email{pelle.ohlsson@acousort.com}
\affiliation{AcouSort AB, Medicon Village, S-223 81 Lund, Sweden}

\author{Henrik Bruus}
\email{bruus@fysik.dtu.dk}
\affiliation{Department of Physics, Technical University of Denmark,\\
DTU Physics Building 309, DK-2800 Kongens Lyngby, Denmark}

\author{Thomas Laurell}
\email{thomas.laurell@bme.lth.se}
\affiliation{Department of Biomedical Engineering, Lund University, Ole R\"{o}mers v\"{a}g 3, 22363 Lund, Sweden}

\date{24 August 2026}

\begin{abstract}
Separation of low-abundance biological objects requires high throughput for practical use of an acoustofluidic system. Increasing the flow rate helps in achieving high-throughput if the acoustic energy density can be increased proportionally, and this may be possible with an efficient coupling of the transducer to the device. In particular, antisymmetric actuation using two electrodes with opposite phases is theoretically proven to enhance the acoustic energy density of the device. In this work, we study the symmetric and antisymmetric actuation mechanisms of an acoustofluidic system using both experiments and three-dimensional numerical simulations. The acoustic focusability experiments show that  under the same electrical input power, the antisymmetric actuation mode performs better than the symmetric actuation, quantified in terms of the normalized width of the band formed by the focused particles. Numerical simulations of this particle bandwidth are performed for both actuation modes, and the results suggest that the antisymmetric actuation mode is more robust than the symmetric one, being weakly dependent of the geometric symmetry properties of the system. The simulation results corroborate the experimental findings, which indicate that the antisymmetric actuation increases the acoustophoretic efficiency and robustness for high-throughput applications.
\end{abstract}

\maketitle


\section{Introduction}
Acoustofluidics has gathered attention in the past decade for its precise, gentle, and contactless handling of biological samples, including separation \cite{Augustsson2012, Grenvall2015, Petersson2018, Ohlsson2018, Karthick2018, Urbansky2019, Olofsson2020}, enrichment \cite{Nordin2012, Antfolk2015, Ohlsson2016, Jakobsson2015, Hammarstrom2012, Richard2019, Broman2021} and washing \cite{Tenje2015} of cells and particles. The combination of microfluidic channels and MHz frequencies creates a strong acoustic radiation force for the manipulation of biological samples. To be relevant for clinical standard procedures, acoustic separation must not only be precise but also accomplished in a sufficient time to uphold short sample-to-answer time. Small sample volumes (less than $1~\SImL$) can be processed sufficiently fast in an acoustofluidic system, whereas larger samples volumes (5-10 mL), necessary when extracting low-abundant biomarkers or preparing cell therapies, becomes excessively time consuming and incompatible with laboratory standards. The two basic ways to actuate acoustofluidic systems are either the use of bulk acoustic waves (BAWs) or surface acoustic waves, where the BAWs typically enables higher volumetric throughput due to the higher acoustic energy density \cite{Urbansky2019}.

When striving for higher flow rates, different strategies have been pursued to compensate for the reduced retention time of cells or particles in the acoustic field. When increasing the flow rate in BAW devices, the acoustic energy density in the microchannel must increase proportionally, which is typically achieved by increasing actuation power to yield the same acoustic focusing outcome. However, at elevated actuation power, the thermal dissipation in the transducer will at some point pose an upper limit, why efficient cooling systems have also been proposed to enable increased throughput at maintained separation performance \cite{Augustsson2012, Adams2012, Urbansky2019}. Alternatively, a longer microchannel can compensate for the increased flow rate by providing longer retention time in the sound field; however, this may increase the size and thus cost of the chip and actuator.

Although not optimal, BAW microsystems have to a large extent been actuated in the transversal mode using a single piezoelectric transducer, often made of lead-zirconate-titanium (PZT), mounted underneath the chip, as depicted in \figref{PZT_mounting_sketch}(a) \cite{Nilsson2004, Lenshof2012, Leibacher2015b, Iranmanesh2015, Karthick2018a, Gautam2018a}. Other approaches to couple the piezo actuator to the chip have been proposed using aluminum plates or wedges that support the coupling of multiple transducers and enable more efficient cooling, as shown in \figref{PZT_mounting_sketch}(b) and  \fignoref{PZT_mounting_sketch}(c) \cite{Adams2012, Grenvall2015, Manneberg2009}. More recent investigations, having the transducer mounted on the side of the chip, see \figref{PZT_mounting_sketch}(d), have demonstrated improved acoustic energy densities in the microchannel when compared to the corresponding transversal mode of operation \cite{Qiu2022}. A critical factor in this case is the glueing of the transducer to the minute contact area at the side of the chip. The benefit is, however, that this configuration supports an antisymmetric mode of actuation as compared to the symmetric chip actuation in the transversal mode. In contrast, perfect symmetric actuation is difficult to realize experimentally due to tolerances in lithography, PZT dicing, and alignment, when gluing the transducer to the chip and placing it on the holder.

\begin{figure}[t]
\centering
\includegraphics[]{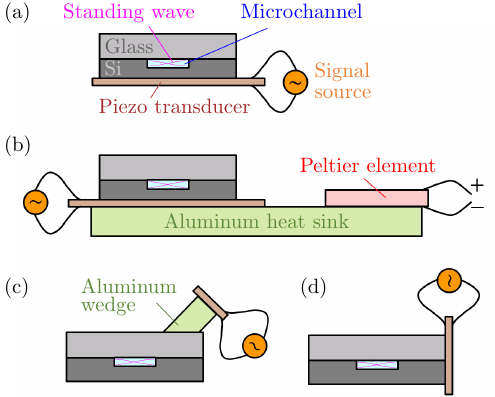}
\caption{\figlab{PZT_mounting_sketch}
Schematic of common piezo transducer coupling strategies in BAW devices: (a) Transversal actuation from the bottom, (b) Transversal actuation  coupled with an aluminum plate for efficient cooling, (c) Transducer coupled with an aluminum wedge at an angle, and (d) Side actuation.}
\end{figure}

Bora \etal were first to demonstrate by numerical simulations, that antisymmetric actuation of a glass chip in a transversal mode configuration, should yield considerably higher acoustic energy density in the microchannel in half-wave resonance mode compared to the traditional symmetric transversal actuation \cite{Bora2015}. In their study, a basic two-dimensional (2D) numerical model without any viscous losses was considered. Later, a more detailed numerical analysis using a basic three-dimensional (3D) model were carried out by Tahmasebipour \etal which further confirmed the finding of the former study \cite{Tahmasebipour2020}. The study also considered geometric symmetry and asymmetry and found via experimental and simulations that the latter produced a higher acoustic energy density inside the square resonators compared to the former design. Moiseyenko and Bruus showed the importance of antisymmetric actuation for polymer-based devices to obtain whole system resonance in the system \cite{Moiseyenko2019}. The principle has been demonstrated later via experimental and numerical study with a cut transducer top for separating the top electrodes into two halves \cite{Lickert2021}. The comparison of symmetric and antisymmetric actuation using the phase difference between the two electrodes for obtaining optimal design and efficiency has, however, not been studied in detail.

In this work, we study the acoustophoretic focusing of suspended particles flowing through a straight microchannel with a transverse standing ultrasound half-wave imposed in a 60-mm-long section of the microchannel. Using the FocuScan method introduced by us in Ref.~\cite{Vitali2019}, we quantify the focusability by the full width at half maximum of the particle distribution across the microchannel width of the particles flowing in a band along the channel after passing through the focusing standing ultrasound wave. This width is termed the particle bandwidth. We experimentally show that antisymmetric actuation improves the focusability and increases the corresponding acoustic energy density in the microchannel compared to symmetric actuation under the same input power to the transducer. We furthermore perform 3D simulations of the experimental device with the symmetric and antisymmetric actuation, and to account for the thin highly damped viscous boundary-layer fields, we use the numerical boundary-layer model BL25 developed recently by us \cite{Hoque2025}. For each design, we validate the simulation results with those of experiments and discuss the physical mechanism of symmetric and antisymmetric actuation. Specifically, we study the effect of geometric asymmetry on the symmetric actuation case, and the discrepancy of the experimental results compared to the previous theoretical studies is elucidated. The numerical simulation predicts
an improved focusability and an increased acoustic energy density when actuating in the antisymmetric mode compared to the symmetric mode, further corroborating our experimental results.

\section{Numerical simulations}
\seclab{num_sim}
The numerical simulation of the experimental device is based on the 3D boundary-layer (BL25) model by Hoque and Bruus \cite{Hoque2025}. It is carried out in COMSOL Multiphysics 6.3 using the \qmarkstt{Weak Form PDE interface} following the earlier works from our group \cite{Hoque2025, Karlsen2016, Joergensen2021, Joergensen2023, Joergensen2023a}.

\subsection{3D device modeling}
\seclab{dev_model}
The 3D model of the device consists of a fluid channel ($\fl$) enclosed in a glass domain ($\sl$), which is actuated in a symmetric or antisymmetric manner using a PZT transducer ($\pz$)  of type Pz26 attached at the bottom via a glue layer ($\mr{gl}$), resembling the actual experimental device. The device, having the dimensions listed in \tabref{param_geom}, is mounted to the microscope platform using a holder, which is modeled by the clamped area ($\mr{cl}$) as shown in \figref{device_sketch}. We study the effects of any unintentional geometric asymmetry in the system using the $y$-offset ($\Dy$) of the channel center to that of the Pz26 center. The PZT transducer is cut at the bottom with a shallow groove (gr) for antisymmetric actuation and is driven by an applied time harmonic AC voltage $\vphn(t) = \frac12 \vphn_0\:\emiot$ with peak-to-peak amplitude $\vphn_0$, angular frequency $\omega = 2\pi f$, and frequency $f$. The AC voltage induces a solid displacement $\uuu_1$, which in turn creates an oscillating acoustic pressure and velocity field $p_1$ and $\vvv_1$ in the fluid. Assuming the actuation voltage to be sufficiently small, the response is well described by first-order time-harmonic perturbation expansions,
\begin{table}[t]
\centering
\caption{\tablab{param_geom} The geometric dimensions of the acoustofluidic device used in the 3D simulations and corresponding to the experimental devices, see \figref{device_sketch}.}
\begin{ruledtabular}
\begin{tabular}{crccr}
 Parameter&  Value &$\qquad$ & Parameter & Value \\ \hline
 $\Lsl$   & $90~\SImm$  &  & $\Lfl  $ & $85~\SImm$\upspace \\
 $\Wsl$   & $2.8~\SImm$ &  & $\Wfl  $ & $430~\SImum$ \\
 $\Hsl$   & $1.4~\SImm$ &  & $\Hfl  $ & $150~\SImum$ \\
 $\Lpz$   & $60~\SImm$  &  & $\Lcl  $ & $10~\SImm$ \\
 $\Wpz$     & $5~\SImm$ &  & $\Wgr$ & $150~\SImum$ \\
 $\Hpz$     & $1~\SImm$ &  & $\Hgr$ & $150~\SImum$ \\
 $\Dy$ & 0-35~$\SImum$  &  & $\Hgl$ & $5~\SImum$ \\
\end{tabular}
\end{ruledtabular}
\end{table}
\begin{figure}[t]
\centering
\includegraphics[width=\columnwidth]{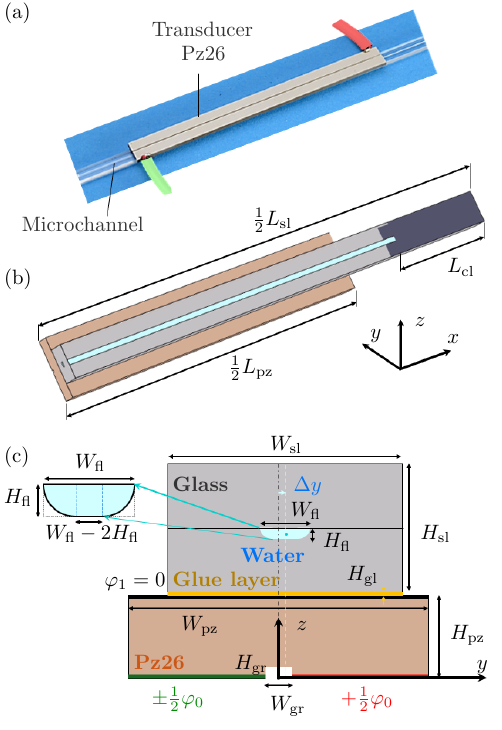}
\caption{\figlab{device_sketch}
(a) An image of the experimental glass device with a straight microchannel attached with a Pz26 transducer from the bottom for actuation. (b) The 3D simulation domain representing half of the experimental device where the top glass wall is not shown for visualizing the fluid channel (light blue) inside the glass (light grey), the transducer attached at the bottom and the clamped area of the device end which is used in the device holder during the experiment. (c) The 2D cross-section of the simulation domain with the split bottom electrode (green and red). $\Delta y$ is the offset of the channel center from that of the piezoelectric transducer.\\[-8mm]}
\end{figure}
 \bsubalat{def_pert_series}{3}
 \eqlab{def_phi} 
 \vphn_1 (\rrr,t) = \vphn_1 (\rrr)\:\emiot,
 \\
 \eqlab{def_p} 
 \uuuI (\rrr,t) = \uuuI (\rrr)\:\emiot,
 \\
 \eqlab{def_v} 
 \pI (\rrr,t) = \pI (\rrr)\:\emiot.
 \esubalat
Using first-order perturbation theory and assuming adiabatic acoustics, such that $\pI =\rhoI \cO^2$ with $\rho_1$ being the density variation, the continuity and Navier--Stokes equation reduces to the Helmholtz equation for the acoustic pressure field $\pI (\rrr)$, with the first-order acoustic velocity $\vvv_1$ in the bulk given by the gradient of the pressure,
 \bsubal{FirstOrderEqu}
 \eqlab{Helmholtz_eqn}
 \Lapl p_1 &= - \Big(1+\ii \Gamma\Big) k_0^2 p_1,\;
 \text{ with }  \Gamma =  \frac{(\frac43\etaO+\etaOb)\omega}{\rhoO \cO^2},
 \\
 \eqlab{1st_order_vel}
 \vvv_1 &=  -\ii\frac{1-\ii\Gamma}{\omega\rho_0}\:\grad \pI.
 \esubal
Here, $\cO$ is the speed of sound in fluid, $\kO = \frac{\omega}{\cO}$ the corresponding wave number, and $\etaO$ and $\etaOb$ is the dynamic and bulk viscosity of the fluid, respectively, see \tabref{param_values}.

\begin{table}[t]
\centering
\caption{\tablab{param_values} Parameters of water and of the solids PZT Pz26, borofloat glass, glue, and polystyrene at $25~\SICel$ used in the numerical simulation. For Pz26 $C_{12} = C_{11} - 2C_{66}$, whereas for glass $C_{12} = C_{11} - 2C_{44}$.}
\begin{ruledtabular}
\begin{tabular}{ccccc}
 Parameter & Value & \hspace*{3mm} & Parameter & Value \\ \hline
 \multicolumn{5}{l}{\emph{Water}~\cite{Muller2014}}\upspace  \\
 $\rhoO$ & $ 997~\SIkgm$ &  & $\etaBO$ & $2.485~\SImPas$ \\
 $\cO$   & $1497~\SImps$ &  & $\Gamma$  & $10.3~\mr{THz}^{-1} f$ \\
 $\kapO$ & $448~\SIpTPa$ &  & $\etaO$	 & $0.890~\SImPas$ \\[1mm]
 \multicolumn{5}{l}{\emph{Bulk lead zirconium titanate}, PZT~\cite{CTS_Ferroperm_Pz26}} \\
 $\rhosl $    & 7700 $\SIkg\:\SIm^{-3}$ &   & $\Gamsl$ & 0.005 \\
 $C_{11}$   & 168 GPa   &  & $C_{33}$	&  123 GPa\\
 $C_{12}$   & 110 GPa   &  & $C_{44}$	& 30.1 GPa\\
 $C_{13}$   & 99.9 GPa  &  & $C_{66}$	& 29.0 GPa\\
 $e_{31}$   &$-2.8~\SIC\:\SIm^{-2}$    &
 & $e_{15}$	& $9.86~\SIC\:\SIm^{-2}$ \\
 $e_{33}$   & 14.7 $\SIC\:\SIm^{-2}$      &
 &  $\Gamma_e$ & $0.02$ \\
 $\ve_{11}$ & 828 $\veO$ &  & $\ve_{33}$ & 700 $\veO$ \\
 $\Gamma_\ve$   & $0.005$ &
 & &
 \\[1mm]
 \multicolumn{5}{l}{\emph{Glass, borofloat}  \cite{SchottD263}} \\
 $\rhosl$ & 2510 $\SIkgm$ &  && \\
 $E$      & 72.9 GPa      &  & $s$  & 0.208           \\
 $C_{11}$ & 81.8 GPa      &  & $C_{44}$    & 30.2 GPa        \\
 $C_{12}$ & 21.5 GPa      &  & $\Gamsl$  &  0.0004  \\
 $\cL$    & 5710 $\SImps$ &  & $\cT$    & 3467 $\SImps$
 \\[1mm]
 \multicolumn{5}{l}{\emph{Glue, NOA 86H}  \cite{Bode2022}} \\
 $\rhosl$ & 1300 $\SIkgm$ &  &&          \\
 $C_{11}$ & (4.65-i0.51) GPa      &  & $C_{44}$    & (1.21-i0.12) GPa        \\
 $\cL$    & 1891 $\SImps$ &  & $\cT$    & 965 $\SImps$
 \\[1mm]
 \multicolumn{5}{l}{\emph{Polystyrene} particles in water  \cite{Barnkob2012a}} \\
 $\rhosl $    & 1050 $\SIkg\:\SIm^{-3}$ &  & $\kappa_\sl$   &  249~$\SIpTPa$  \\
 $f_0$ & 0.444  &   & $a$ & 2.5~$\SImum$ \\
 $f_1$ & 0.034  &   & $\Phi$   &  0.17    \\
\end{tabular}
\end{ruledtabular}
\end{table}

We further assume that all the solid domains consist of a linear elastic material with density $\rhosl$, and in addition the piezoelectric transducer has a linear dielectric response and zero charge density. The equation of motion of the displacement field $\uuu_1 (\rrr)$ {and the electrical potential $\vphn_1 (\rrr)$ in the transducer is therefore given by the Cauchy and Gauss equation, respectively,
\bsubal{transducerEq}
-\rhosl \omega^2\: \uun_1 &= \div \sigmabf,\\
\zerovec = \div \DDD &= \div (-\vebf \cdot \grad \vphn_1),
\esubal
where $\DDD$ is the electrical displacement field and $\vebf$ is the dielectric tensor. The first-order solid velocity $\VVV_1$ is given by the time harmonic displacement field $\uuuI$,
\beq{1st_order_wall_vel}
 \VVV_1 =  -\ii\omega \uuuI.
\eeq
In the Voigt notation, the electromechanical coupling for the piezoelectric material PZT Pz26 is given by
 \bal
 \sigma_{1xx}&= \Cnn_{11}\pp_x\unn_{1x} + \Cnn_{12}\pp_y\unn_{1y} + \Cnn_{13}\pp_z\unn_{1z}+e^{{}}_{31}\pp_z \vphn_1,
 \nn \\
 \sigma_{1yy}&= \Cnn_{12}\pp_x\unn_{1x} + \Cnn_{11}\pp_y\unn_{1y} + \Cnn_{13}\pp_z\unn_{1z}+e^{{}}_{31}\pp_z \vphn_1,
 \nn \\
 \sigma_{1zz}&= \Cnn_{13}\pp_x\unn_{1x} + \Cnn_{13}\pp_y\unn_{1y} + \Cnn_{33}\pp_z\unn_{1z}+e^{{}}_{33}\pp_z \vphn_1,
 \nn \\
 \sigma_{1yz}&= \Cnn_{44}(\pp_y \unn_{1z}+\pp_z \unn_{1y}) + e^{{}}_{15}\pp_y \vphn_1,
 \nn \\
 \sigma_{1xz}&= \Cnn_{44}(\pp_x \unn_{1z}+\pp_z \unn_{1x}) + e^{{}}_{15}\pp_x \vphn_1,
 \nn \\
 \eqlab{StressStrainPiezo}
 \sigma_{1xy}&= \Cnn_{66}(\pp_x \unn_{1y}+\pp_y \unn_{1x}),
 \\
 D^{{}}_x &= e^{{}}_{15}(\pp_x \unn_{1z} + \pp_z \unn_{1x}) - \ve^{{}}_{11}\pp_x \vphn_1,
 \nn \\
 D^{{}}_y &= e^{{}}_{15}(\pp_y \unn_{1z} + \pp_z \unn_{1y}) - \ve^{{}}_{11}\pp_y \vphn_1,
 \nn \\ \nn
 D^{{}}_z &= e^{{}}_{31}\pp_x\unn_{1x} + e^{{}}_{31}\pp_y\unn_{1y} + e^{{}}_{33}\pp_z\unn_{1z}- \ve^{{}}_{33}\pp_z \vphn_1.
 \eal
The remaining three components of the stress tensor are given by the symmetry relation $\sigma_{ik} = \sigma_{ki}$. The values of the coefficients are listed in \tabref{param_values}.

Similarly, the Cauchy equation~\eqnoref{transducerEq} governs $\uuu_1$ in the purely elastic solids, but now the stress-strain relation~\eqnoref{StressStrainPiezo} includes only the first six equations, as $\DDD$ and $\varphi$ are zero and do not couple to $\sigmabfsl$ and $\uuu_1$. These parameter values are also listed in \tabref{param_values}.

\subsection{Electrical admittance and power dissipation}
A potential difference $\vphn = \vphn_0$ is applied between the bottom grounded electrode and the top electrodes as either symmetric or antisymmetric actuation. The electrical admittance $Y$ is obtained by the ratio of the surface integral of the current density $-\ii\omega\DDD\cdot\nnn$ to $\vphn_0$ \cite{Lickert2021},
 \beq{impedanceSim}
 Y =  \frac{I}{\vphn_0},
 \;\text{ with }\;
 I = -\ii\omega\int_{\pp\Omega} \DDD\cdot\nnn \:\dm a.
 \eeq
Here, the integral is performed over the bottom electrode surface. Using \eqref{impedanceSim}, the electrical power dissipation $\Pin$ in the PZT transducer averaged over one oscillation period is given by
 \beq{avgPowerDiss}
 \Pin = \langle \vphn_0 I \rangle
 = \frac12 \re\big[|\varphi_0|^2 \; Y \big]
 = \frac12 |\varphi_0|^2 \; \re\big[Y\big].
 \eeq

\subsection{Acoustic energy density and radiation force}
\seclab{EacFrad}
The average acoustic energy density $\Eac$ of the fluid in the volume $V_\mr{fl}$ and acoustic radiation force $\FFFrad$ on a particle of radius $a$ in a dilute suspension in the fluid with compressibility $\kapO$ and scattering coefficients $\fO$ and $\fI$, are computed from the first-order fields as~\cite{Lickert2021},
 \bsubal{EacFrad}
 \eqlab{EacDef}
 \Eac &=
 \frac{1}{V_\mr{fl}}\:\int_{V_\mr{fl}} \Big[\frac14\kapO |\pI|^2 + \frac14\rhoO|\vvvI|^2 \Big]\:\dm V,
 \\
 \eqlab{FradDef}
 \FFFrad &= \pi a^3\grad\Big[\frac13 \fO \kapO |\pI|^2  - \frac12 \fI \frac14\rhoO|\vvvI|^2 \Big].
 \esubal

\subsection{Boundary conditions}
\seclab{bc_1st-order}
In this section, we summarize the boundary conditions of the first-order fields in the solid boundaries and fluid-solid interfaces. We assume that the surrounding air leads to zero stress on all outer solid surfaces and zero free charge density on the surface of the PZT transducer. For the AC potential $\vphn_1$, we assume the top electrode to be grounded, and the bottom electrodes to be supplied with a voltage of the same phase for symmetric \qmarks{sa} and a 180-degree phase shift for the antisymmetric \qmarks{asa} actuation,
 \bsubalat{BCelec3D}{2}
 \vphn_1 &= 0,  && \text{top grounded},
 \\
 \vphn_1 &= +\tfrac12 \vphn_0,+\tfrac12 \vphn_0, \quad &&\text{bottom in-phase for \qmarks{sa}},
 \\
 \vphn_1 &= -\tfrac12 \vphn_0,+\tfrac12 \vphn_0, \quad && \text{bottom anti-phase for \qmarks{asa}}.
 \esubalat

If the actuation is at constant voltage, the actuation potential is simply $\vph_0^\text{cv} = \vphn_0$. However, if the actuation is at constant power $\Pin$, the actuation potential is re-scaled as $\vph_0^{\text{cp}} = K(f) \vph_0$, where $K(f)$ is a frequency-dependent scaling factor chosen to ensure the given constant dissipated power $\Pin$ \cite{Lickert2022},
\beq{scale_param_cp}
K(f) = \sqrt{\frac{2\:\Pin}{\varphi_0^2 \; \re\big[Y(f)\big]}}.
\eeq
On the internal solid interfaces between the PZT-electrode, the electrode-glue layer and the glue layer-glass, the stress and solid displacement fields are continuous. Similarly, at the fluid-solid interfaces the stress and velocity in the solid and in the fluid are continuous, which we ensure by applying the recently developed boundary-layer model BL25 \cite{Hoque2025}, where the viscous boundary layers are taken into account analytically as modifications to the boundary conditions for the bulk acoustic fields at the fluid-solid interface. Thus, we do not need to resolve the very thin boundary layers of thickness $\delta\: \sim 0.5\:\SIum$, and it is feasible to do a complete 3D simulation. The first-order velocity and stress continuity conditions are
 \bsubal{BL25_1st}
 \eqlab{p1_bc}
 \ppperp{p_1} &=\; \frac{\ii\omega\rho_0}{1-\ii\Gamma} \big(\vwallperp-\frac{\ii}{\ks}\pardiv \vvvwallpar \big)-\frac{\ii}{\ks}\big(\kc^2p_1+\ppperpsqr p_1\big),
 \\
 \eqlab{stress1_bc}
 \sigmabf^{\mr{sl}}_1 \cdot \een_\zs &= -p_1\ezs\; +\;
 \ii\ks \eta_0 \big[\vvvwall_1-\vvv_1 + (\vwallperp-v_{1\zs})\een_\zs\big].
 \esubal
Here, $\een_\zs$ is the unit normal vector at the solid-fluid interface  pointing in the perpendicular $\zs$-direction from the solid into the fluid, $\parallel$ represents the in-plane components, $\vvvwall_1 = -\ii \omega \uuu_1$ with $\vvvwall_1 = \vvvwallpar+\vwallperp \een_\zs$ is the solid wall velocity, $\kc = \kO (1+\ii\frac12\Gamma)$ is the compression wave number, and $\ks = (1+\ii)/\delta$ is the shear wave number.

In contrast to our previous work, we include a model of the chip holder in the form of clamping areas on the top and bottom outer device surfaces near the ends of the device, represented by the dark gray regions in \figref{device_sketch}(b). The clamping is modeled as Dirichlet boundary condition on the solid displacement field,
 \beq{clamped_bc}
 \uuu_1 = \zerovec,\; \text{ on the clamped part of the surface}.
 \eeq
Finally, by assuming the $yz$-plane to be a symmetry plane, the computational domain is reduced to half of the full domain. The above boundary conditions thus need to be supplemented by the following symmetry-plane conditions valid for each perturbation order,
 \bsubalat{BC_symmetry}{2}
 \partial_{x}\varphi &= 0,\;  &
 \pp_x \pI &= 0,
 \\
 u_{1,x} &= 0,\; &
 \sigma_{yx}^{\mr{sl}}= \sigma_{zx}^{\mr{sl}} &= 0,\quad
 \\ \nn
  \text{}&\text{on the}& \text{ symmetry plane } x=0. &
 \esubalat
In summary, the governing equations \eqsnoref{Helmholtz_eqn}{transducerEq} are solved with the boundary conditions \eqnoref{BCelec3D} for the transducer, the BL25 model \eqnoref{BL25_1st} for the solid-fluid boundary, the clamping \eqnoref{clamped_bc} for the solid, \eqnoref{BC_symmetry} for the symmetry plane, as well as stress- and charge-free outer surfaces.

\begin{figure}[t]
\centering
\includegraphics[width=\columnwidth]{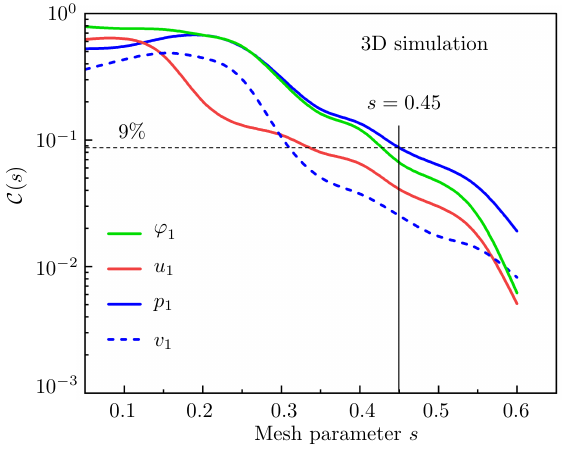}
\caption{\figlab{3D_mesh_conv}
Semilog mesh convergence plot of the convergence parameter versus the scaling mesh parameter for the first-order fields. The mesh for the simulation is considered at $s = 0.45$ with a maximum discrepancy of 9~\%.}
\end{figure}

\subsection{Mesh convergence}
\seclab{mesh_conv}
We use a 3D tetrahedral mesh to discretize the fluid, solid and PZT domains. A mapped mesh is used for the glue layer between the Pz26 and the glass domain. A given mesh is characterized by the mesh-scale parameter $s$, and the mesh convergence is carried out using the convergence parameter $\calC[a(s)]$ for any field variable $a(s)$~\cite{Muller2012},
 \beq{err_conv} 
 \calC[a(s)] = \sqrt{\frac{\int_\Omega \mid a(s)-a(s_{\text{max}}) \mid^2 \dm V}{\int_\Omega \mid a(s_{\text{max}}) \mid^2 \dm V}}.
 \eeq
The initial rough mesh with $s=0.05$ consists of 10,187 tetrahedral domain elements, 3,170 boundary elements and 1,041 edge elements. The corresponding degrees of freedom for the solver are 161,969, and the required computational time is 22~$\SIs$.

In \figref{3D_mesh_conv} is shown how $\calC[a(s)]$ evolves for $a = \vph_1, u_1, p_1, v_1$ as $s$ varies from $s=0.05$ (a rough mesh) to $s=0.65$, a fine mesh. As the mesh is refined, the computational time increases significantly. At the mesh scale $s = 0.45$, the resolved mesh consists of 154,464 tetrahedral domain elements, 17,008 boundary elements and 1,989 edge elements. The corresponding degrees of freedom for the solver are 2,237,826. The estimated error on the computed fields is in the range 3-8\%, and the computational time required for a single frequency is about 12~min. Since the computational time increases significantly with the mesh parameter, we used the mesh corresponding to $s = 0.45$ for all the simulations. The corresponding maximum mesh sizes in the solid and in the fluid channels are $h_{\text{max}}^{\mr{sl}} = 333~\SIum$ and $h_{\text{max}}^{\fl} = 83~\SIum$, respectively. All simulations were carried out on workstations of the type HP Z4 G5 Intel(R) Xeon(R) w3-2525@3.50 GHz processor with 512 gigabytes of random access memory and 64-bit operating system.

\subsection{Simulated focused particle bandwidth}
\seclab{focusability_sim}

As described in \secref{FocuScan}, the experimental device focusability is quantified in terms of the width of the focused particle band using the FocuScan method~\cite{Vitali2019}. To enable a direct comparison of this quantity with simulation, we have developed the following algorithm based on our earlier work on particle focusing \cite{Barnkob2010, Barnkob2012}: We compute the pressure field $\pI$ with average acoustic density $\Eac$ in the fluid channel induced by a given actuation voltage amplitude. We then use the COMSOL \qmarkstt{general projection operator} to project $\pI(x,y,z)$ in the channel section above the transducer of length $\Lpz = 60~\SImm$ along the $x$-direction onto the $yz$-plane. The projected pressure field turns out to be independent of not only $x$ but also of $z$, especially near resonance, so we fit it to the standard one-dimensional (1D) sinusoidal standing pressure wave $\hat{p}_1(y)$ with acoustic energy density $\Eac$,
 \beq{p1hat}
 \hat{p}_1(y) = \hat{p}_a \sin(\hat{k}y),\; \text{ with }\;
 \Eac = \frac14 \kapO \hat{p}^2_a,
 \eeq
where the wave number $\hat{k}$ and the pressure amplitude $\hat{p}_a$ are fitting parameters. Next follows a 3D steady state simulation of the axial flow profile $v_x(y,z)$ in the $yz$ cross section of the straight channel with its amplitude adjusted to fit the experimental flow rate $Q_\mr{exp} = 150~\SImuL/\SImin$. The velocity values $v_x(y_i,z_j)$ were subsequently obtained on a regular grid $\big\{y_i,z_j\big\} = \big\{(i-50.5)\Delta\Wfl, (j-14.0)\Delta\Hfl\big\}$ in the $yz$ plane for $\Delta\Wfl = \Wfl/100$, $\Delta\Hfl = \Hfl/30$, $1\leq i \leq 100$, and $1 \leq j \leq 27$, but only including the $N_p = 2276$ grid points  inside the quarter-circle rounded channel cross section of area $A_\mr{ch} = \big[\Wfl+(\frac12\pi-2)\Hfl\big]\Hfl = 0.85 \Wfl\Hfl$ shown in the inset of \figref{device_sketch}(c). The velocity grid is then extracted as an excel file to be processed in a Matlab script \qmarkstt{NPB} together with $\hat{p}_1(y)$ to compute the normalized particle bandwidth (NPB) of the focused particles.

The \qmarkstt{NPB} Matlab script consists of nine steps. Here,
Step (1) defines the axial velocity field $v_x(y,z)$ for any position $(y,z)$ using the Matlab interpolation function \qmarkstt{interp2} based on the grid data $\big\{y_i,z_j, v_x(y_i,z_j)\big\}$.

Step (2) makes an ensemble of $N_p$ spherical polystyrene particles of radius $a = 2.5~\SImum$ with initial center positions $\big\{y_i^0, z_j^0\big\} = \big\{y_i+\delta y_i, z_j + \delta z_j\big\}$, where $\delta y_i$ and $\delta z_j$ are random offsets drawn from random uniform distributions on the respective intervals $[-1, 1]\times\frac12 \Delta\Wfl$ and $[-1, 1]\times\frac12 \Delta\Hfl$ using the Matlab random function \qmarkstt{rand}.

Step (3) computes for each particle $ij$ the 3D trajectory $\rrr_{ij}(t) = \big\{x_{ij}(t), y_i(t), z_j(t)\big\}$ at time $t$ as follows. From the starting position $\rrr_{ij}(0) = \big\{0, y_i^0, z_j^0\big\}$ at time zero in the vertical cross section at $x = 0$ above the upstream edge of the PZT transducer, a time-integration is performed on the quasi-steady equation of motion $\FFFrad\big[\rrr_{ij}(t)\big] + \FFFdrag\big[\rrr_{ij}(t)\big] = \zerovec$, which expresses force balance between the Stokes drag force $\FFFdrag\big[\rrr_{ij}(t)\big] = 6\pi\eta a \big[ v_x\big(y_i(t),z_j(t)\big)\:\een_x - \pp_t\rrr_{ij}(t)\big]$ and the radiation force $\FFFrad$, \eqref{FradDef}, computed at position $\rrr_{ij}(t)$. The special form~\eqnoref{p1hat} of the pressure $\hat{p}_1(y)$ combined with the corresponding acoustic velocity~\eqnoref{1st_order_vel}, implies according to \eqref{FradDef} that the radiation force $\FFFrad$ acts only in the $y$-direction. This specific situation has been analyzed in Refs.~\cite{Barnkob2010, Barnkob2012}, where the particle position $\big\{y_i(t), z_j(t)\big\}$ in the channel cross section is found to be~\footnote{Note the coordinate systems: here $-\frac12 \Wfl < y < \frac12 \Wfl$, whereas in Refs.~\cite{Barnkob2010, Barnkob2012} $0 < y < \Wfl$}
 \bsubal{yDynamics}
 \eqlab{yt_pos}
 y_i(t) &= \frac{1}{\hat{k}}\bigg\{\text{sign}(y_i^0)\frac{\pi}{2}
 +\arctan\!\Big[\!\tan\Big(\frac{\pi}{2}\!+\!\hat{k} y_i^0\Big)\ee^{t/\tac}\!\Big]\bigg\},
 \\
 \eqlab{zt_pos}
 z_j(t) &= z_j(0),
 \\
 \eqlab{tac}
 \tac &= \frac{3\etaO}{4 \Phi \hat{k}^2 a^2 \Eacfl} = \frac{3\etaO}{\Phi \hat{k}^2 a^2 \kapO \hat{p}^2_1},
 \esubal
with $\tac$ being the acoustophoretic time scale and $\Phi$ the usual acoustic contrast factor of a single particle in the fluid.  The equation of motion of $x_{ij}$ is simply $\partial_t x_{ij} = v_x\big[y_i(t),z_j(0)\big]$, which is integrated numerically using the Matlab ordinary-differential-equation solver \qmarkstt{ode45},
 \beq{xt_pos}
 x_{ij}(t) = \int_0^t v_x\big[y_i(t'),z_j(0)\big]\:\dm t'.
 \eeq

Step (4) in \qmarkstt{NPB} Matlab script computes the transit time $t_{ij}^{\Lpz}$ it takes a particle to be advected by the flow field $v_x(y,z)$ through the acoustically active region from $\big\{0, y_i^0, z_j^0\big\}$ above the upstream edge to $\big\{\Lpz,y_i^{\upLpz}, z_j^{\upLpz}\big\} = \big\{\Lpz,y_i\big(t_{ij}^{\Lpz}\big), z_j^0\big\}$ above the downstream edge of the transducer. Given \eqref{xt_pos}, $t_{ij}^{\Lpz}$ must fulfil the equation
 \beq{tiL_def}
 \Lpz = \int_0^{t_{ij}^{\Lpz}} v_x\big[y_i(t'),z_j(0)\big]\:\dm t',
 \eeq
which is solved for $t_{ij}^{\Lpz}$ by taking a discrete number of times $t_k$, with $k = 1,2, \ldots , N_t$, and forming the data set $\big\{t_k,x_{ij}(t_k)\big\}$. The value $\big\{t_{ij}^{\upLpz},\Lpz\big\}$ is found by interpolation using the Matlab function \qmarkstt{interp1} based on this data set.

In step (5), the final position $\big\{x_{ij}^{\upLpz},  y_i^{\upLpz}, z_j^{\upLpz}\big\}$ and velocity $v_{x,ij}^{\upLpz}$ in the vertical cross section at $x = \Lpz$ of each particle in the ensemble,  are computed from \eqsref{yDynamics}{tiL_def},
 \bsubal{final_ij}
 \eqlab{yf_pos}
 \big\{x_{ij}^{\upLpz},  y_i^{\upLpz}, z_j^{\upLpz}\big\} &= \big\{\Lpz, y_i\big(t_{ij}^{\upLpz}\big), z_j^0\big\},
 \\
 v_{x,ij}^{\upLpz} &= v_x\big(y_i^{\upLpz}, z_j^0 \big).
 \esubal

In step (6), the particles in a given ensemble are sorted at their final position $x = \Lpz$ in a histogram containing $\Nbin = 49$ bins $B_q$, $q = 1, 2, \ldots, \Nbin$, of width $\Wbin = (y_\mr{max}-y_\mr{min})/\Nbin \approx 8.1~\SIum$, across the microchannel in the $y$-direction from $y = y_\mr{min} = -198~\SImum$ to $y = y_\mr{max} = +198~\SImum$ ($34~\SImum$ less then the full channel width to avoid edge effects), as follows:
 \beq{ijINq}
 \begin{array}{c}
 \text{Particle $ij$ is in bin $B_q$ (written as '$ij \in B_q$') if}
 \\[2mm]
 q = \mr{floor}\bigg[\frac{y_i^{\upLpz}-y_\mr{min}}{\Wbin} \bigg] + 1.
 \end{array}
 \eeq

In step (7), the number $N_q$ of particles in each bin $B_q$ is computed using the steady-state condition $\pp_t n = 0$ of the local particle density $n(x,y,z,t)$. For $x < 0$ before the PZT transducer, the uniform distribution of particles with the typical volume fraction $\phi = 0.001$ results in the initial uniform and constant density $n^0 = \phi/V_p$, where $V_p$ is the volume of a single particle. In the initial cross section plane at $x = 0$, the small average area available per grid point $ij$ is $\Delta A_{ij}^0 = \Delta \Wfl \Delta \Hfl$, and thus the particle current $\Delta \dot{n}^0$ through this area centered at $\big(0, y_i^0, z_j^0 \big)$ is given by $\Delta \dot{n}^0 = n^0 v_{ij}^0 \Delta A_{ij}^0$. Due to the particle focusing, these incoming particles end up passing the final cross section plane at $x = \Lpz$ through the deformed area $\Delta A_{ij}^{\upLpz}$ centered around $\big(\Lpz, y_i^{\upLpz}, z_j^{\upLpz} \big)$ with velocity $v_{ij}^\upLpz$, and local density $n_{ij}^\upLpz$. Thus the particle current there is $\Delta \dot{n}^\upLpz = n_{ij}^\upLpz v_{ij}^\upLpz \Delta A_{ij}^\upLpz$. The steady-state condition implies a constant particle current, so $\Delta \dot{n}^0 = \Delta \dot{n}^\upLpz$, and thus we find the particle density
 \beq{nijLpz}
 n_{ij}^\upLpz = n^0 \frac{v_{ij}^0 \Delta A_{ij}^0}{v_{ij}^\upLpz \Delta A_{ij}^\upLpz}.
 \eeq
The number of particles $N_q$ that are in bin $B_q$ and in the region $\Lpz < x < \Lpz+\Lbin$ right after the PZT transducer is computed as
 \beq{Nq}
 N_q = \sum_{ij\in B_q} n_{ij}^\upLpz \Lbin \Delta A_{ij}^\upLpz =  n^0  \Lbin \Delta A_{ij}^0 \sum_{ij\in B_q} \frac{v_{ij}^0}{v_{ij}^\upLpz}.
 \eeq
Finally, choosing $\Lbin$ such that the total number of particles in the above region equals the number $N_p$ of initial particles (grid points), and using $n^0 = \phi/V_p$ as well as $A_{ij}^0 = \Delta \Wfl \Delta \Hfl$, we arrive at the final expression for the simulated number of particles $N_q$ in bin $B_q$
 \beq{NqFinal}
 N_q = \phi \frac{\Lbin \Delta \Wfl \Delta \Hfl}{V_p} \sum_{ij\in B_q} \frac{v_{ij}^0}{v_{ij}^\upLpz},
 \;\text{ with }\; N_p = \sum_{q=1}^{\Nbin} N_q.
 \eeq

In step (8), we repeat steps (1) - (7) nine times and thus obtain 10 statistically independent particle histograms. We compute the average histogram and the standard deviation of $N_q$ in each bin $B_q$.

\begin{figure}[t]
\centering
\includegraphics[width=\columnwidth]{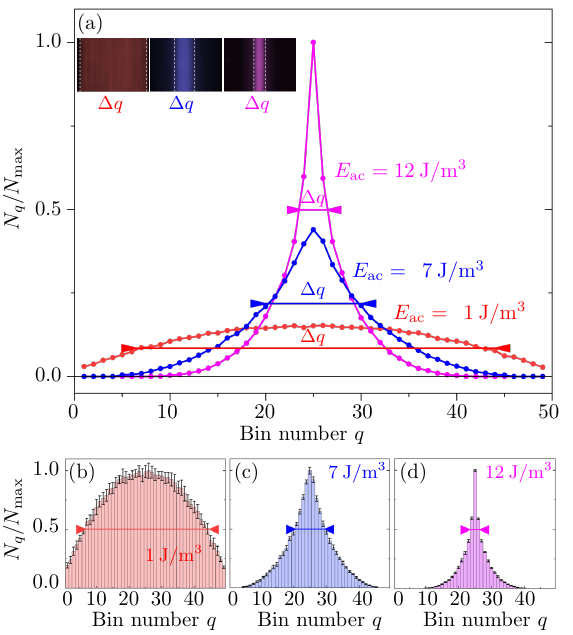}

\mbox{}\\[-7mm]
\caption{\figlab{histograms}
(a) For three values of the acoustic energy density [$\Eac = 1$ (red), 7 (blue), and $12~\SIJ/\SIm^3$ (magenta)] in the antisymmetrically actuated device, the normalized particle number histogram $N_q/N_\mr{max}$ is plotted versus bin number $q$ spanning the horizontal width of the microchannel of the final particle positions $y_i^\upLpz$ after acoustophoretic focusing along the active part of length $\Lpz = 60~\SImm$ at a flow rate of 150~$\SImum$L/min. Each histogram contains $\Nbin = 49$ bins each of width $\Wbin = 8.1~\SImum$ and is the average of 10 ensembles each containing $N_p = 2276$ particles. The normalization is chosen to be $N_\mr{max} = 428$, which is the maximum number of particles in the bins for $\Eac =  12~\SIJ/\SIm^3$. The horizontal lines terminated by arrow heads represent the full width at half maximum.
(b) The detailed histogram normalized by  $N_\mr{max} = \mr{max}_q\{N_q\} = 65$ for $\Eac = 1~\SIJ/\SIm^3$ and with one-standard-deviation error bars from the ensemble average. Almost no focusing is obtained, and $\beta_\mr{sim}= 0.77$.
(c) Same as panel (b), but for $\Eac = 7~\SIJ/\SIm^3$ and with $N_\mr{max} = 186$. Fairly strong focusing is obtained, and $\beta_\mr{sim}= 0.19$.
(d) Same as panel (c), but for $\Eac = 12~\SIJ/\SIm^3$ and with $N_\mr{max} = 428$. Very strong focusing is obtained, and $\beta_\mr{sim}= 0.06$.
}
\end{figure}

Finally,  in step (9) of  the \qmarkstt{NPB} Matlab script, we obtain the normalized particle bandwidth from the average histogram. We define the two interpolation functions $N^\notop_<(q)$ and $N^\notop_>(q)$ using \qmarkstt{interp1} based on the data sets $\big\{q,N_q\big\}_{q\leq25}$ and $\big\{q,N_q\big\}_{q\geq25}$, respectively, and then compute the non-integer bin-numbers $q^\notop_<$ and $q^\notop_>$ by demanding $N^\notop_<(q^\notop_<) = \frac12 N_\mr{max}$ and $N^\notop_>(q^\notop_>) = \frac12 N_\mr{max}$, where $N_\mr{max} = \mr{max}_q\{N_q\}$ is the number of particles in the bin with most particles in the histogram. The value $\beta_\mr{sim}$ for the normalized particle bandwidth obtained by simulation is given by
 \beq{NPB_sim}
 \beta_\mr{sim} =  \left\{\begin{array}{cl}
 \frac{1}{q_\mr{max}}\:(q^\notop_> - q^\notop_<), & \text{ for }\: N_\mr{min} < \frac12 N_\mr{max},
 \\[1mm]
 0.95, &  \text{ for }\: N_\mr{min} > \frac12 N_\mr{max},
 \end{array} \right.
 \eeq
where $N_\mr{max}$ and $N_\mr{min}$ is the maximum and minimum number of particles in the bins of the average histogram, and the number 0.95 takes into account the experimentally observed depletion of particles near the walls. Examples of simulated particles distribution histograms and normalized particle bandwidths $\beta_\mr{sim}$ are shown in \figref{histograms}

\begin{figure*}[t]
\centering
\includegraphics[width=\textwidth]{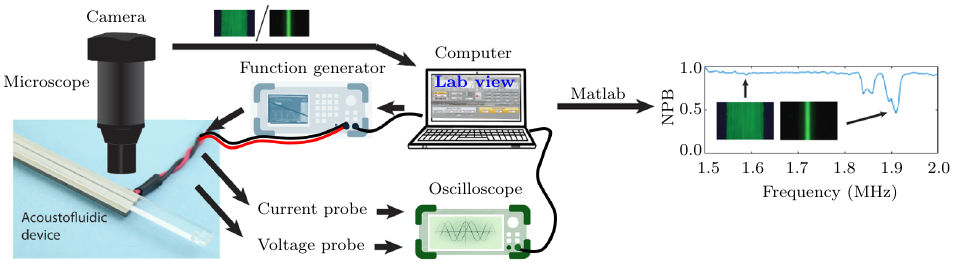}
\caption{\figlab{focuscan_sketch}
Schematic sketch of the FocuScan method. Function generator, current probe, voltage probe, oscilloscope and microscope camera were controlled by LabVIEW. Images of fluorescent beads flowing through the channel were collected at the end of separation channel while performing a frequency sweep. Post-processing in Matlab generated a plot of the normalized particle bandwidth (NPB) versus the frequency.}
\end{figure*}

\section{Experimental procedure}
\seclab{exp_proc}

\subsection{Microfluidic chip}
\seclab{exp_chip}
Acoustofluidic chips were designed in collaboration with AcouSort (Lund, Sweden) and manufactured by Micronit Microfluidics (Enschede, Netherlands) in borosilicate glass by photolithography and wet etch. The device consisted of two sandblasted inlets and outlets connected by a 7-$\SIcm$ long, 430-$\SImum$ wide, and 150-$\SImum$ high, isotropically etched, separation channel. Fluidic connections were enabled by clamping the chip into a holder (AcouSort, Lund, Sweden), with connected tubing. The setup is sketched in \figsref{device_sketch}{focuscan_sketch}.

\subsection{Acoustic actuation}
\seclab{acoustic_act}
Three microfluidic chips, all with the same design, were used for comparison in this study. They were actuated by piezoelectric transducers (Pz26, CTS Ferroperm Piezoceramics, Denmark), $60\times5\times1$~mm,  glued underneath the glass chip. To enable antisymmetric and symmetric actuation, the transducers had a 150~$\SIum$ wide groove along the centreline of the transducer to generate two separated electrodes on one side of the transducers. The transducers were glued underneath the chip with the groove facing away from the chip, see \figref{device_sketch}.
In experiments, the transducer was driven with a 180-degree phase shift between the electrodes separated by the groove to generate antisymmetric actuation. The phase shift was enabled by connecting signal and ground to each of the divided top electrodes [red and green in \figref{device_sketch}(c)]. The bottom electrode [black in \figref{device_sketch}(c)] facing the chip was left floating. When driving the electrodes without a phase shift, the signal source was connected to both top electrodes while ground was connected to the bottom electrode. In this way symmetric actuation was enabled. The microfluidic chips were actuated to generate a half-wavelength standing wave along the width of the channel, generating a pressure node in the center of the channel.

\subsection{FocuScan setup and focusability experiments}
\seclab{FocuScan}
FocuScan, an in-house developed LabVIEW program (National Instruments Corporation, USA), enabled automatic control of the function generator (Tektronix, AFG3022B, UK), oscilloscope (Tektronix, TBS 2000B, UK) and microscope camera (Infinity 1-2, Lumenera, Canada) \cite{Vitali2019}. The acoustofluidic device was placed under a stereo microscope (Olympus, SZX16, Japan) and the microscope camera was used to visualize the end part of the separation channel from above. Only one inlet and one outlet were used during these experiments, the other inlet and outlet were plugged. A syringe pump (neMESYS, Cetoni GmbH, Germany) controlled the flow (150~\SImuLpmin) of fluorescent beads (4.9~$\SIum$ in diameter, Fluoro-Max, Thermo Fisher Scientific, MA, USA) through the separation channel. FocuScan performs a frequency sweep and for each frequency, 10 images are collected and post processed in MATLAB (2019b, MathWorks) to calculate the average normalized particle bandwidth $\beta$, \eqref{NPB_sim}. When particles occupy the full width of the channel $\beta = 1$, whereas $\beta$ drops as the beads gets acoustically focused to the pressure node along the centreline when operated at a resonance as shown in \figref{focuscan_sketch}. FocuScan was also further developed to include a voltage probe (Tektronix, TPP0100, UK) and a current probe (Tektronix, CT-2, UK) to enable frequency sweeps with constant power applied.

\subsection{Acoustic energy density measurement}
\seclab{Eac_measurement}
Particle image velocimetry (PIV) experiments were performed using a confocal microscope (Eclipse Ti2, Nikon, Japan) equipped with a spinning disk unit (X-Light V3, Crest, Italy) and a CMOS camera (Prime 95B, Teledyne Photometrics, 33AZ), combined with a laser diode emitting blue fluorescent excitation light with a peak wavelength of 488 nm, and a fluorescence filter cube with excitation passband from 475 to 495 nm and emission passband from 510 to 531 nm. A 10$\times$ objective lens and 0.3 numerical aperture provided a (1.83$\times$1.83)~mm$^2$ field of view. The focal plane was placed in the mid-height of the channel, and the optical slice thickness was 5.7 $\SIum$, thus excluding particles close to bottom and top were not captured. PIV measurements were performed for one chip, for both symmetric and antisymmetric actuation. Measurements were performed for the 60-mm-long part of the channel covered by the piezoelectric transducer. With a 1.83-mm-long field of view, 33 sections were recorded to cover the region of interest. At each section, a minimum of 3 repeats were performed. For a few sections with high acoustic energy and fast focusing, 4 or 5 repeats were made to ensure that a sufficient number of frames were collected. For all sections, 50-440 frames were collected. The piezoelectric transducer was driven by a function generator (33220, Agilent Technologies, Inc., CA, USA) and a PicoScope (5244D, Pico Technology, UK) monitored the applied voltage as well as the resulting current to ensure an input power of 12.5 mW to the transducer. Fluorescent green particles (Fluoro-Max, Thermo Fisher Scientific, MA, USA) with a diameter of 4.9 $\SIum$ were used in PIV experiments. An open source PIV algorithm, PIVlab (MATLAB), was used to process the data \cite{Thielicke2014}.

%
%

\section{Results and discussion}
\seclab{results_discussion}

\mbox{}\\[-18mm]
\subsection{FocuScan sweep and optimal frequency range}
\seclab{opt_freq}

\mbox{}\\[-6mm]
The FocuScan software (see \secref{FocuScan}) was used to find the resonance frequency by searching for the minimum NPB in a frequency sweep \cite{Vitali2019}. We studied the acoustofluidic device named C operating at constant voltage $\vph_0^{\text{cv}}$ first to obtain the resonance frequency range from the experiment. The experiments were done with a peak-to-peak voltage amplitude of 1.5~$\SIV$ for the antisymmetric actuation. However, for the symmetric actuation, it was ramped up to $\vph_0^{\text{cv}} = 3\: \SIV$ to better observe the weaker particle focusing. First, a wide-range frequency sweep was made between 1.5-2.0~$\SIMHz$ in 5-$\SIkHz$ increments to find regions of interest  (resonances) for the acoustofluidic chip. The normalized particle bandwidth is plotted versus frequency for the symmetric (\qmarkstt{sa}) and antisymmetric (\qmarkstt{asa}) actuation in \figref{focusability_cv}(a) and (b), respectively. Even with 3.0-V actuation voltage in the symmetric case, the focusability is better in the case of antisymmetric actuation with a lower 1.5-V actuation voltage.

In simulations, the frequency is varied from 1.5 to 1.8~$\SIMHz$ in increments of 10~$\SIkHz$, whereas in the region of interest from 1.8 to 2.0~$\SIMHz$, the increments are reduced to 4~$\SIkHz$. Treating the channel-center offset $\Delta y$ shown in \figref{device_sketch} as a fitting parameter, the best fit of the simulation to the experimental data was obtained by $\Dy = 80~\SIum$. For this offset, the simulated NPB $\beta_\mr{sim}$ for both actuation symmetries is plotted versus frequency in \figref{focusability_cv} together with the experimental data, and we see a fair agreement between the two in the optimal frequency range from 1.82 to  2~MHz (the gray-shaded area): For \qmarkstt{sa} in panel (a) the two barely resolved experimental double-dips at $\fres = 1.840/1.860$ and $1.895/1.910~\SIMHz$ (solid blue arrows) are matched in simulation with a relative deviation of less than 1~\% by the double-dips at $\fres = 1.856/1.868$ and $1.952/1.910~\SIMHz$ (shaded blue arrows). Similarly for \qmarkstt{asa} in panel (b) the two experimental dips at $\fres = 1.840$ and $1.910~\SIMHz$ (solid red arrows) are matched in simulation with a relative deviation of about 2~\% by the dips at $\fres = 1.880$ and $1.956~\SIMHz$ (shaded red arrows).

\begin{figure}[t]
\centering
\includegraphics[width=0.9\columnwidth]{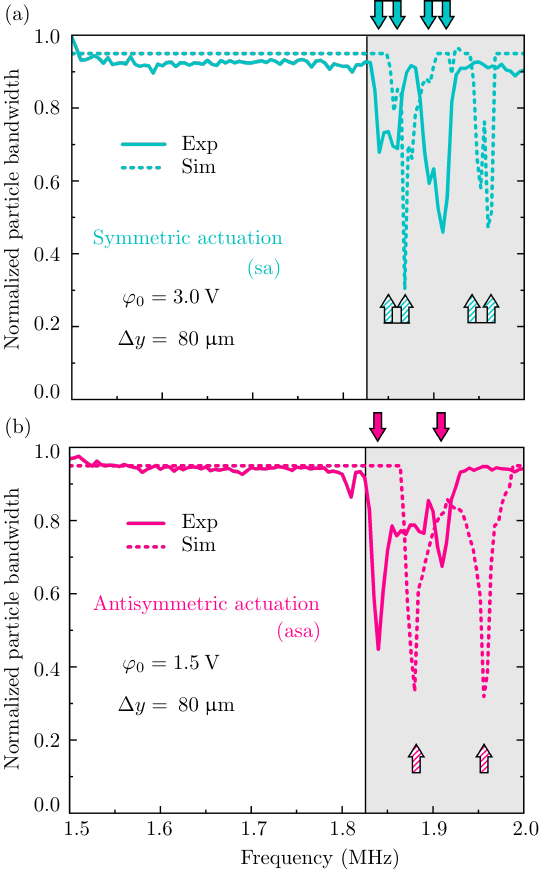}

\caption{\figlab{focusability_cv}
Plots of FocuScan sweeps of the normalized particle bandwidth $\beta$ versus frequency at a fixed flow rate of $Q = 150~\SImuL/\SImin$ from experiment and corresponding simulation for device C with a channel offset $\Delta y = 80~\SImum$.
(a) Symmetric actuation "\sa" at 3~$\SIV_\mr{pp}$ for
\underline{ex}p\underline{eriment} (solid blue lines) with resonance double-dips at $\fres = 1.840/1.860$ and $1.895/1.910~\SIMHz$ (filled blue arrows), and for \underline{simulation} (dashed blue lines) with double-dips at $\fres = 1.856/1.868$ and $1.952/1.910~\SIMHz$ (shaded blue arrows).
(b) Antisymmetric actuation "\asa" at 1.5~$\SIV_\mr{pp}$ for
\underline{ex}p\underline{eriment} (solid red lines) with resonance dips at $\fres = 1.840$ and $1.910~\SIMHz$ (filled red arrows), and for
\underline{simulation} (dashed red lines) with dips at $\fres = 1.880$ and $1.956~\SIMHz$ (shaded red arrows).
}
\end{figure}

\begin{figure}[t]
\centering
\includegraphics[width=0.9\columnwidth]{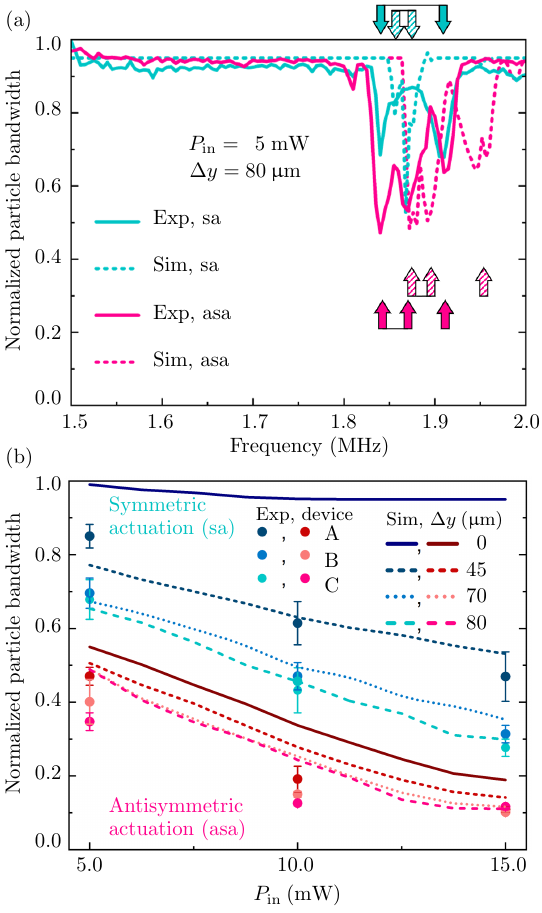}
\caption{\figlab{NPB_cp}
(a) Normalized particle bandwidth $\beta$ for a device having a fixed flow rate  $Q = 150~\SImuL/\SImin$ and a channel offset $\Delta y = 80~\SImum$ with symmetric \qmarks{sa} and antisymmetric \qmarks{asa} actuation at constant power $\Pin = 5~\SImW$.
\underline{Ex}p\underline{eriment}:
\qmarks{sa} (solid blue lines) with resonance dips at $\fres = 1.840$ and $1.910~\SIMHz$ (filled blue arrows) and
\qmarks{asa} (solid red lines) with resonance dips at $\fres = 1.840$, 1.875, and $1.910~\SIMHz$ (filled red arrows).
\underline{Simulation}:
\qmarks{sa} (dotted blue lines) with dips at $\fres = 1.856$ and $1.868~\SIMHz$ (shaded blue arrow) and
\qmarks{asa} (dotted red lines) with dips at $\fres = 1.872$, 1.892, and $1.948~\SIMHz$ (shaded red arrows).
(b) Mean value of $\beta$ versus input power $\Pin$ with error bars obtained from a 5-fold repetition of each measurement at the resonance frequency. \underline{Ex}p\underline{eriment}: Device A, B, and C, each at $\Pin = 5$, $10$, and 15~mW for both \qmarks{sa} (blue points) and \qmarks{asa} (red points). \underline{Simulation}: $\beta_\mr{sim}$ versus $\Pin$ for channel-center offset $\Delta y = 0$, 45, 70, and $80~\SImum$ for both \qmarks{sa} (blue lines) \qmarks{asa} (red lines), where the three latter are chosen to fit the data points.
}
\end{figure}

\subsection{Acoustic focusability}
After finding the optimal frequency range, we study the focusability at the resonance frequency versus the electrical input power $\Pin$ for three nominally identical devices labeled A, B, and C. This allowed for comparing the two actuation methods on multiple devices without performing time consuming PIV measurements.

We measured the normalized particle bandwidth $\beta$ for symmetric and antisymmetric actuation for the three devices at $\Pin = 5.0$~mW as shown for device C in \figref{NPB_cp}(a). Then, we measured $\beta$ at the obtained resonance frequency for the three input power levels $\Pin = 5$, 10, and $15~\SImW$ for each device. Each of these nine particle focusing experiments were repeated five times to obtain both a mean value and a standard deviation on the obtained $\beta$ vs.\ $\Pin$ data points, which is shown in \figref{NPB_cp}(b). Subsequently, we simulated each of the three devices for nine values of $\Pin$ in the range from 5 to 15~mW. We computed $\beta_\mr{sim}$ of \eqref{NPB_sim} and used the experimental data to fit channel-center displacement $\Delta y$ of each device. As shown in \figref{NPB_cp}(b), we found the best fit to be $\Delta y = 45$, 70, and $80~\SImum$ for device A, B, and C, respectively.

Returning to \figref{NPB_cp}(a) for a more detailed discussion of device C, we note that the measured $\beta$ vs.\ frequency for constant input power $\Pin$ appears much like the results for constant voltage actuation shown in \figref{focusability_cv}. Both in experiment and simulation it is seen in \figref{NPB_cp}(a) that the focusability is better for antisymmetric actuation compared to the symmetric actuation. At $\Pin = 5$~mW, the minimum NPB was found in experiments to be 0.47 for antisymmetric and only 0.68 for symmetric actuation, and the corresponding values were found in simulation to be 0.46 and 0.53. In the experiments for symmetric actuation (solid blue line), two distinct resonance dips are observed at $1.840$ and  $1.910~\SIMHz$ (filled blue arrows). These are matched in simulation by the two dips at at $1.856$ and  $1.860~\SIMHz$ (shaded blue arrows) differing from experiment by 1\% and $-3$\%, respectively. Similarly, for the antisymmetric actuation (solid red line), three resonance dips are observed at $1.840$, 1.875, and  $1.910~\SIMHz$ (filled red arrows), which are matched in simulation by the three dips at at $1.872$, 1.892, and $1.948~\SIMHz$ (shaded blue arrows) differing from experiment by only 2\%, $-1$\%, and 2\%, respectively.

On \figref{NPB_cp}(b), the experimental values for $\beta$ are plotted as colored points (bluish for \qmarks{sa} and reddish for \qmarks{asa}) with standard-deviation error bars obtained from a the 5-fold repetition of each measurement. It is noticeable, that the variation in the results of each device at a given input power is relatively low, although the repeated experiments were carried out at different days, whereas a larger and systematic variation is observed between the nominally identical devices. As mentioned above, we ascribe this variation to differences in the dicing and mounting of piezoelectric transducer, and we model it by the channel-center offset $\Dy$ in the simulations chosen for each device to fit the experimental data. In all measured and simulated cases, $\beta$ at the resonance frequency is lower for the antisymmetric actuation than for symmetric actuation. This confirms that antisymmetric actuation generates higher acoustic energy density, in agreement with previous studies \cite{Bora2015, Moiseyenko2019, Tahmasebipour2020}. Our simulations further confirm that $\beta$ is nearly independent of the channel offset $\Dy$ for antisymmetric actuation, but a small improvement in the focusability can be seen as $\Dy$ is increased from zero, whereas $\beta$ is strongly dependent on $\Dy$ for symmetric actuation, as the focusability improves from zero ($\beta = 1$) as $\Dy$ increases from zero. These findings are in good agreement with the experimental data that show a significantly larger inter-device variability of $\beta$ when operated with symmetric actuation as compared to antisymmetric.

Now, turning to the variation of the input power also shown in \figref{NPB_cp}(b), we find experimentally that the particle focusing for antisymmetric actuation is already good with $\beta \approx 0.4$ at $\Pin = 5~\SImW$, nearly complete with $\beta \approx 0.2$ at $\Pin = 10~\SImW$, and practically complete with $\beta \approx 0.1$ at 15~$\SImW$. This behavior is well captured by the simulation, which also seems to indicate that the focusability is slightly worse for an ideal device with perfect geometrical symmetry, as $\beta_\mr{sim} \approx 0.55, 0.35$, and 0.2 for the three experimental power levels, respectively. For the symmetric actuation, the best but not very good focusability is obtained for device C ($\Delta y = 80~\SImum$), closely followed by device B ($\Delta y = 70~\SImum$), with $\beta \approx 0.69, 0.43$, and 0.28 for the three power levels, respectively. Device A has a worse performance with  $\beta \approx 0.85, 0.62$, and 0.48. Although the antisymmetric actuation outperforms the symmetric actuation, the later still exhibits some focusability, even at low powers ($<10\:\SImW$). The introduction of the channel-center offset $\Dy$ in the simulation seems to explain why the symmetric actuation does induce some focusability: The offset breaks the geometrical symmetry of the device and thus introduces a non-zero antisymmetric component in the actuation. We speculate that similarly in a real device, any minor asymmetry in the fabricated device geometry or transducer mounting will contribute to some antisymmetric actuation even for a symmetrically applied actuation voltage. This aspect is studied further in the next section.

\subsection{Acoustic energy density}
To validate the focusability results, PIV measurements of the particle focusing velocity $\vvn_p$ were carried out along the full length of the channel in the 60-mm long segment covered by the transducer, see \figref{Eac_exp}(a). From these measurements, the local acoustic energy density $\Eac(x)$ along the channel was deduced from averaging $\vvn_p$ in 33 segments each 1.8~mm long, see \figref{Eac_exp}(b). Due to the time consuming procedure, this measurement was only carried out for one the devices at resonance with symmetric and antisymmetric actuation for the constant input power $\Pin = 12.5$~mW. Nevertheless, this constitutes an adequate validation, since the focusability results indicate that the relative energy density between antisymmetric and symmetric actuation is nearly the same for all three of the tested acoustofluidic chips. The average acoustic energy density $\Eac = \avr{\Eac(x)}$ for the antisymmetric actuation is $\Eac^\asa = 10.4~\SIJpmc$, which is 4.8 times larger than for the symmetric actuation, $\Eac^\sa = 2.2~\SIJpmc$. Moreover, $\Eac(x)$ exhibits a stronger a more narrow maximum peak or \qmarks{hot spot} for \qmarks{asa} than for \qmarks{sa} as shown in \figref{Eac_exp}(b). We speculate that the observed spatial variation of $\Eac(x)$ along the length of the channel may be attributed to a combination of axial modes and the non-uniformity of the glue-layer thickness in mounting the PZT transducer onto the chip.

\begin{figure}[t]
\centering
\includegraphics[width=\columnwidth]{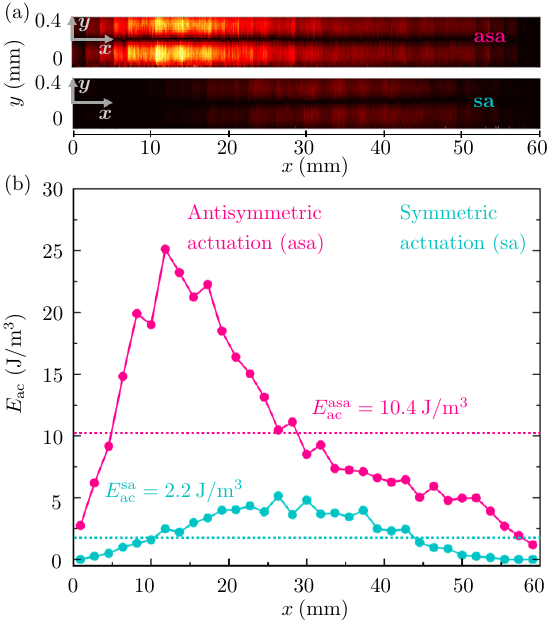}
\caption{\figlab{Eac_exp}
(a) Color plots of the the $xy$-dependence of the magnitude $\vnn_p$ of the particle focusing velocity from 0 (black) to $240~\SImumps$ (white) measured for one device by PIV in the active region in the microchannel above the 60-mm long PZT transducer. The results are shown for antisymmetric (asa) and symmetric (sa) actuation both with a constant power input $\Pin = 12.5$~mW. (b) The $x$-dependence of the acoustic energy density $\Eac(x)$ for both \qmarks{asa} and \qmarks{sa} deduced the average $\vnn_p$ in 33 segments each 1.8~mm long along the channel.}
\end{figure}

The geometric asymmetry plays an important role in coupling acoustic energy into the fluid filled microchannel, especially in devices with symmetric actuation. To better understand the phenomenon we performed simulations to obtain the acoustic energy density $\Eac$ of the acoustofluidic device as a function of the channel-center offset $\Dy$ for symmetric and antisymmetric actuation at the resonance frequency, both with constant power $\Pin = 12.5 \: \SImW$. The results of this study are presented in \figref{Eac_sim}(a) for the fixed resonance frequency $\fres = 1.876~\SIMHz$. This frequency was determined, in analogy with \figref{NPB_cp}, as the resonance dips in the $\beta$-versus-frequency curve for the fixed channel-center offset $\Dy = 80~\SIum$ and input power $\Pin = 12.5~\SImW$, and it was found to be the same for actuation modes. Not surprisingly, both actuation modes in \figref{Eac_sim}(a) show increasing $\Eac$ as $\Dy$ increases from zero to $100~\SImum$, since $\Dy$ induces more antisymmetry the larger it is. But the two actuation modes differ in one crucial aspect: The \qmarks{asa}-mode is already antisymmetric for $\Dy = 0$, so $\Eac^\asa$ increases linearly from 8 to $11~\SIJ/\SIm^3$, a variation of only 14\% around the mean value $9.5~\SIJ/\SIm^3$. In contrast, the \qmarks{sa}-mode starts out being fully symmetric at $\Dy = 0$, so $\Eac^\asa$ increases linearly from  0 to $7~\SIJ/\SIm^3$, a variation of 100\% around its mean value $3.5~\SIJ/\SIm^3$. From this we conclude that the \qmarks{asa}-mode is more controllable and less dependent of geometric uncertainties than the \qmarks{sa}-mode, but also that the \qmarks{sa}-mode can induce a large $\Eac$, if the device has a pronounced geometrical asymmetry, such as placing the chip at one edge of a wide PZT transducer as has been done in many previous studies of acoustic particle focusing.

As shown in \figref{NPB_cp}(b) the experimentally measured normalized particle bandwidth agreed well with the simulated one for a channel-center offset $\Dy = 80~\SIum$. For this particular value we therefore also computed $\Eac$ for \qmarks{asa}- and the \qmarks{sa}-mode and marked them by red and blue circles, respectively, in \figref{Eac_sim}.

The simulated values are $\Eac^\asa =  10.7~\SIJpmc$ and $\Eac^\sa = 6.0~\SIJpmc$, respectively, which are 3~\% and 173~\% higher than the experimental values 10.4 and 2.2~$\SIJpmc$ found in \figref{Eac_exp}(b). It is interesting to note that in the numerical simulations, the spatial variations in $\Eac(x)$ are much less pronounced than in the experimental results shown in \figref{Eac_exp}(b). This corroborates our speculation that the actual variations in $\Eac(x)$ is induced by inhomogeneities in the actual devices not taken into account in the simulation model. That fact that the numerical simulation only deviates from the measured $\Eac$ by 3\% corroborates the observation made above that the \qmarks{asa}-mode is less dependent of such inhomogeneities. In contrast, the more sensitive \qmarks{sa}-mode is harder to predict accurately in the simulation.

\begin{figure}[t]
\centering
\includegraphics[width=\columnwidth]{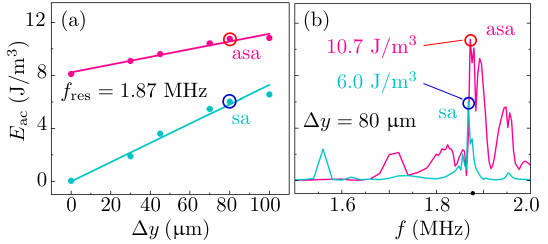}
\caption{\figlab{Eac_sim}
The simulated acoustic energy density $\Eac$ for constant input power $P_\mr{in} = 12.5~\SImW$ for antisymmetric (asa, red points and spline curves) and symmetric (sa, blue points and spline curves) actuation. (a) Plots of $\Eac$ versus center-axis offset $\Delta y$ for $f=1.876~\SIMHz$. (b) Plots of $\Eac$ versus frequency $f$ for $\Delta y = 80~\SImum$. In  both panels the encircled points correspond to $f=1.876~\SIMHz$ and $\Delta y = 80~\SImum$.}
\end{figure}

\begin{figure}[t]
\centering
\includegraphics[width=0.95\columnwidth]{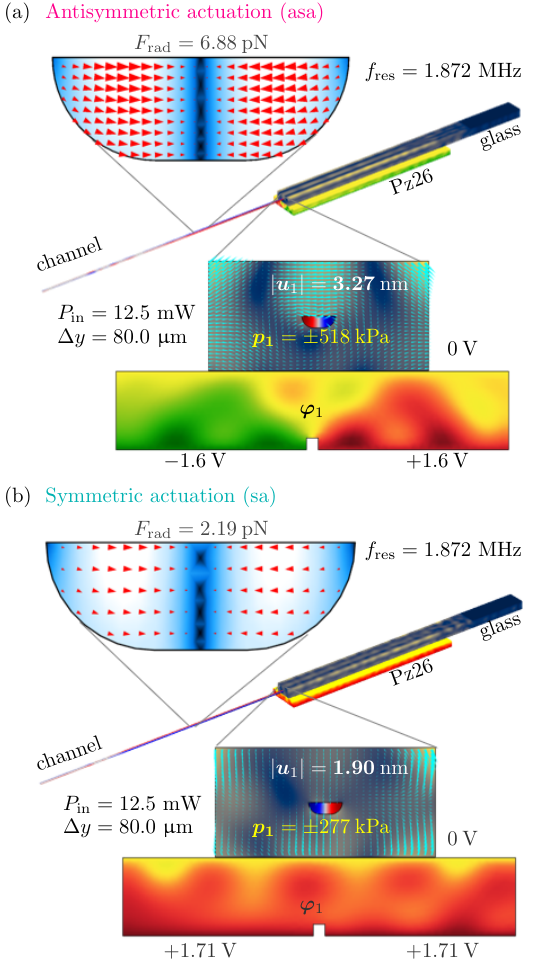}
\mbox{}\\[-4mm]
\caption{\figlab{fields_sim} Plots of the simulated first-order fields in 3D and vertical 2D cross sections at the resonance $f=1.876~\SIMHz$ identified in \figref{Eac_sim} with center-axis offset $\Delta y = 80~\SImum $ and input power $P_\mr{in} = 12.5~\SImW$ for (a) the antisymmetric actuation (asa) and for (b) the symmetric actuation (sa). The amplitude of the color plot for each field is listed. The direction of the radiation force $\FFFrad$ is shown by red arrows.\\[-8mm]}
\end{figure}

Finally, in \figref{fields_sim}, we show plots of the various fields obtained by simulation at the resonance frequency $f = 1.876~\SIMHz$ at input power $\Pin = 12.5~\SImW$ for both actuation symmetries in a device with a channel-center offset $\Dy = 80~\SIum$. The solid displacement is higher (3.27~nm) and with more deformation happening at the top of the microchannel for the antisymmetric actuation compared to the symmetric actuation (1.90~nm). Thus, a stronger antisymmetric mode is created in the former case, which leads to a pressure half-wave with amplitude 378~$\SIkPa$, 1.7 times higher than that of symmetric actuation. We also calculated the acoustic radiation force inside the fluid channel using \eqref{FradDef} for each actuation mode. We find $\Frad = 6.44~\SIpN$ and $\Frad = 2.98~\SIpN$ for the antisymmetric and symmetric actuation respectively, a ratio of 2.2 times, which implies a 2.2 times faster focusing with antisymmetric actuation.

\section{Conclusion}
\seclab{Conclusion}
By experiments and numerical simulations, we have demonstrated the importance of the actuation mechanism via symmetric and antisymmetric actuation to enhance the focusing ability of an acoustofluidic device for high-throughput applications. The quantitative comparison of the two actuation symmetries was obtained using the normalized particle bandwidth $\beta$, which was measured using the FocuScan method~\cite{Vitali2019} and simulated using the BL2025 model~\cite{Hoque2025} combined with an efficient algorithm for computing particle bandwidths, which supplements the traditional simulation of acoustic energy densities. The simulations agreed well with the experiments, typically within a few percent for the resonance frequencies and about 20\% for $\beta$ and the acoustic energy density $\Eac$.

A main result is the demonstration that the geometric asymmetry of the device is less critical to $\beta$ and $\Eac$ for antisymmetric actuation, and this robust configuration allows for good focusability and efficient energy conversion to the fluid. The symmetric actuation was shown to be more sensitive to unintentional geometric asymmetries, and generally leads to a reduced focusability and a smaller $\Eac$.

To better understand the effects of geometric asymmetries, simulations were carried out in which the channel-center offset $\Dy$ was increased from 0 to $100~\SImum$, and it was shown that the measured differences in $\beta$ between three nominally identical devices, could be modeled by choosing a different value of $\Dy$ for each device. For both actuation symmetries, the acoustic energy density $\Eac$ increases linearly with $\Dy$, but less prominently for the antisymmetric actuation, since it already has a high $\Eac$ for $\Dy = 0$, whereas the symmetric actuation increases from $\Eac = 0$ for $\Dy = 0$.  In general, the numerical simulations confirmed our experimental observations, and further, they emphasized the importance of antisymmetric actuation as a more robust configuration for efficient energy transfer to the fluid channel. Our results also show that device designs aimed at being completely symmetric are hard to realize experimentally due to unintentional symmetry breaking during the device fabrication.

Future work can be done for the improvement of devices with antisymmetric actuation by optimizing the design of piezoelectric transducers, coupling layer, and the geometry of the elastic solid. Additionally, one can use thin-film transducers instead of bulk transducers, which in simulations performed well with more controlled position and robust coupling with the solid device \cite{Steckel2021b}.

\section*{Acknowledgements}
This work was supported
by the \textit{BioWings} project funded by the European Union's Horizon 2020 \textit{Future and Emerging Technologies} (FET) programme, grant No.~801267,
by the \textit{ACOUSOME} project funded by the European Innovation Council (EIC), \textit{HORIZON EIC 2022 TRANSITION}, grant No.~101099787, and by the Swedish Research Council grant No.~2019-00795. K.A. and S.Z.H. contributed equally to this work.

%
%


%

\end{document}